\documentclass[pra,bibnotes,twocolumn]{revtex4}
\usepackage{graphicx}
\begin{document}
\draft
\def\ds{\displaystyle}
\title{PT Symmetric Floquet Topological Phase }
\author{C. Yuce}
\address{Department of Physics, Anadolu University, Turkey }
\email{cyuce@anadolu.edu.tr}
\date{\today}
\begin{abstract}
In this paper, we study the existence of Floquet topological insulators for $\mathcal{P}\mathcal{T}$ symmetric non-Hermitian Hamiltonians. We consider an array of waveguide in 1D with periodically changing non-Hermitian potential and predict the existence of Floquet topological insulators in the system. We also extend the concept of Floquet topological phase to a two dimensional non-Hermitian system.
\end{abstract}
\maketitle

\section{Introduction}

Recent discovery of topological insulators has attracted great interest \cite{hasan} (reference therein). A topological insulator has gapped energy spectrum in the bulk while it has gapless robust edge states. The concept of topological insulator has recently been generalized to periodically driven systems \cite{FTI1,FTI2,FTI3,FTI4,FTI5,FTI6}. Driving a topologically trivial system into a non-trivial topological phase is possible by a periodic driving. The topological insulators for periodically driven systems are known as Floquet topological insulator (FTI). Many different groups are searching for the existence of both topological and Floquet topological insulators in photonics context \cite{f0,f1,f2,f3,f4,f5,f6,f7,f8,f9}. Haldane and Raghu introduced the photonic analogue of topological insulator in 2008 \cite{f0}. Photonic Floquet topological insulators were first experimentally realized by assuming that the periodicity in time is replaced by a periodicity in propagating direction \cite{f1}. We emphasize that there are some differences between the description of standard and Floquet topological insulators. For example, the Chern number, which provides the net number of chiral edge states crossing the band gap, is not good enough to give a full characterization of the topological properties of periodically driven systems. It was shown in  \cite{anomolus} that so-called anomalous edge phenomena appears only for FTI.\\
The existence of topological phases has recently been investigated for non-Hermitian Hamiltonians \cite{PTop2,PTop3,PTop4,hensch,PTop1,ekl56,cemyuce}. Of special importance is $\mathcal{P}\mathcal{T}$ symmetric non-Hermitian Hamiltonians, where $\mathcal{P}$ and $\mathcal{T}$ operators are parity and time reversal operators, respectively. This is because of the fact that Hermiticity requirement was shown to be replaced by the analogous condition of $\mathcal{PT}$ symmetry for the reality of the spectrum \cite{bender2}. It is well known that a $\mathcal{PT}$ symmetric non-Hermitian Hamiltonian admits real spectrum as long as non-Hermitian degree is below than a critical number. The search for topological phase was first considered by Hu and Hughes \cite{PTop2} and Esaki et al.  \cite{PTop3}. They found that topological states are not stable because of the existence of complex energy eigenvalues. In \cite{PTop2}, they considered Dirac-type non-Hermitian Hamiltonians and concluded that the appearance of the complex eigenvalues is an indication of the non-existence of the topological insulator phase in non-Hermitian models. Similarly, topological phase was shown to be unstable for a non-Hermitian generalizations of the Luttinger Hamiltonian and Kane-Mele model \cite{PTop3}, a one dimensional non-Hermitian tight binding model \cite{hensch} and non-Hermitian Su-Schrieffer-Heeger (SSH) model with two conjugated imaginary potential located at the edges of the system \cite{PTop1}. Recently, we have found stable topological phase for a non-Hermitian $\mathcal{PT}$ symmetric system for the first time in the literature \cite{cemyuce}.\\
The existence of topological phase transition for $\mathcal{PT}$ symmetric non-Hermitian systems is an important question. In the literature, FTI for a non-Hermitian system has not been studied yet. In this paper, we study the existence of FTI for some non-Hermitian Hamiltonians. We first consider 1D SSH model with periodically changing non-Hermitian potential and show the existence of FTI. Secondly, we study a 2D system and predict FTI in the 2D system.

\section{Floquet Topological Insulator}

Consider a non-Hermitian periodical Hamiltonian in the propagation direction with the period $T$, $\ds{H(z)=H(z+T)}$. The energy spectrum of a periodical Hamiltonian can be found using the Floquet theory \cite{cem}. The Floquet spectrum is  $2\pi/T$ periodic in quasi energy just as a repeated zone scheme in quasi momentum for conventional band structure. According to the Floquet theorem, there exist solutions to the Schrodinger equation of the form $\psi_{\alpha}=e^{-i\epsilon_{\alpha}{t}}\Phi_{\alpha}$, where $\Phi_{\alpha}$ is the periodical Floquet eigenstate with period $T$ and the time-independent eigenvalues $\ds{\epsilon_{\alpha}}$ are the Floquet quasi energies. Note that the quasi energies are constants in time. Let us define the Floquet Hamiltonian as $\ds{H_F=H-i \partial/\partial_z}$ that satisfies $\ds{H_F\Phi_{\alpha}=\epsilon\Phi_{\alpha}}$. Since the Floquet states are periodical, they satisfy the relation $\Phi_{\alpha n}=\Phi_{\alpha}e^{in\omega z}$, where the orthonormality condition reads $<<\Phi_{\alpha n}|\Phi_{\beta m}>>=\frac{1}{T}\int_0^Tdz<\Phi_{\alpha n}|\Phi_{\beta m}>=\delta_{\alpha\beta}\delta_{mn}$ and the second bra-ket notation here is used to denote the integration over the propagation direction $z$. Of particular interest for a non-Hermitian periodical Hamiltonian is the situation when all quasi energies are real. In the present paper, we numerically find quasi energy spectrum for our non-Hermitian Hamiltonian. \\
Consider now  a 1D tight binding waveguide array. Let the operators $\ds{a^{\dagger}_n}$ and $\ds{a_n}$ create and annihilate a particle on site  $\ds{n}$, respectively. We are interested in the system with staggered coupling coefficients (i.e. alternate between weak and strong coupling coefficient). Let $\ds{t_1}$ and $\ds{t_2}$ denote alternating coupling coefficients. This model, known as SSH model \cite{SSH}, is a two band model exhibiting nontrivial topological properties. In addition to the SSH model, suppose periodically changing non-Hermitian impurities are inserted in the whole lattice. Let the parameter $\gamma$ denotes the non-Hermitian degree of the impurities. The impurities with $\gamma>0$ and $\gamma<0$ represent gain and loss materials, respectively. The Hamiltonian for our system reads
\begin{eqnarray}\label{mcabjs4}
H=-\sum_{n}t_na^{\dagger}_{n} a_{n+1}+h.c.+(-1)^ni\gamma \cos(\omega z)  a^{\dagger}_n a_n
\end{eqnarray}
where the position dependent coupling coefficients $t_n=t_1$ when $n$ is odd and $t_n=t_2$ when $n$ is even and $\omega$ is the angular frequency of the modulation. This Hamiltonian describes the propagation of an optical field in a tight binding waveguide array with periodically changing gain and loss in the propagation direction. If we replace the propagation direction $z$ by time $t$, then the same Hamiltonian describes a tight binding bichromatic optical lattice with time dependent gain/loss terms.\\
We can rewrite the Hamiltonian as $\ds{H=\sum_{n,m}  a^{\dagger}_n H_{nm}a^{\dagger}_m}$, where $H_{nm}$ are the corresponding matrix elements. Provided the system has translational symmetry (the system is subjected to periodic boundary conditions), the matrix form of the Hamiltonian can then be written in momentum space as
\begin{equation}\label{mczclaz}
H(k)=t_2\sin(2ka)\sigma_y-\left(t_1+t_2\cos(2ka) \right)\sigma_x+i\gamma\cos(\omega z)\sigma_z
\end{equation}
where $\vec{\sigma}$ are Pauli matrices, the crystal momentum $k$ runs over the first Brillouin zone, $-\pi/a<k<\pi/a$ and $\ds{a}$ is the lattice spacing. Below we will show that the system supports topologically nontrivial edge states if $\omega\neq0$. Let us define $t_1=1-\lambda\cos{\Phi}$ and $t_2=1+\lambda\cos{\Phi}$, where $\lambda<1$ is a real valued constant and the phase parameter $\Phi$ is an additional degree of freedom. To study topological phase in our system, we vary $\Phi$ from $0$ to $2 \pi$ continuously and see if the band gap closes during the continuous change. Let us first analyze the case with $\omega=0$. In this case, the corresponding band structure can be readily found. It is given by $\ds{E=\mp  \sqrt{t_1^2+t_2^2+2t_1t_2\cos(2ka)- \gamma^2} }$. As can be seen, the system has two bands. Consider first Hermitian limit, $\gamma=0$. In this case, the band structure is gapped when $\Phi=0$. The band gap decreases by varying $\Phi$ and vanishes at $\Phi=\pi/2$. If we further increase $\Phi$, the band gap opens again. As a result, topological phase transition occurs at $\Phi=\pi/2$ (also at $\Phi=3\pi/2$). Consider now the presence of non-Hermitian term in the Hamiltonian, $\ds{\gamma\neq0}$. In  this case, topological states are not stable since complex energy eigenvalues appear when the phase transition occurs. This can easily be seen by setting $t_1=t_2$ and $ka=\pi/2$ on the above energy expression. As a result, we say that no stable topological phase exists in our non-Hermitian system when $\omega=0$. Let us now search for the existence of Floquet topological phase by considering periodically modulated non-Hermitian impurities, $\ds{\omega\neq0}$. In this case, we apply Floquet theory to find the quasi energy spectrum. In practice, a truncated Floquet Hamiltonian is used in our numerical computation. The corresponding matrix is a $2(2N_F+1)\times2(2N_F+1)$ matrix, where $N_F$ is large enough that the result doesn't depend on $N_F$ \cite{hanggi}. Strikingly, we numerically find stable topological phase provided oscillation frequency $\omega$ is large and the non-Hermitian degree $\gamma$ is below than a critical value. In the Fig.1, we plot the quasi energy spectrum as a function of $\ds{ka}$ when $\omega=2\pi$ and $\gamma=1$. Note that the spectrum is $\ds{\omega}$-periodic in quasi energy and $\ds{2\pi/a}$-periodic in crystal momentum $k$. Just as in the Hermitian case, topological phase transitions can take place by varying the coupling coefficients as long as generic symmetries of the Hamiltonian are preserved. Suppose $t_2$ is adiabatically changed from $0.5t_1$ to $1.5t_1$, where $t_1=1$. As can be seen from the figure, the conduction and valence bands are separated by a gap when $t_2<t_1$. Quantum phase transition occurs at $t_2=t_1$ since the band gap closes at $ka=\mp\pi/2$. If we further increase $t_2$ such that $t_2>t_1$ the band gap opens again. Two periodically driven systems are defined to be topologically equivalent if the quasi-energy band structures can be smoothly deformed from one to the other without closing the band gap. So, we say that the two systems with $t_2<t_1$ and $t_2>t_1$ are topologically distinct since the quasi-energy gap doesn't stay open during the continuous deformation that respects the symmetries of the Hamiltonian. Our non-Hermitian system is topologically trivial when $t_1>t_2$ and nontrivial when $t_1<t_2$. We would like to emphasize that the energy spectrum is always real during the continuous deformation. So the system is topologically stable as opposed to the non driven system with $\omega=0$. The topological distinction between the two cases $t_1<t_2$ and $t_1>t_2$ leads to the presence of topologically protected edge states for the former case. More precisely, both cases are insulators in the bulk, but the case with $t_1<t_2$ possess topologically protected edge states. To see the edge states, the non-trivial topological insulator can be placed in a vacuum, which is topologically equivalent to a conventional insulator. Let us now study topological edge states for a system with $N$ sites. We adopt open boundary conditions with $n = 1$ and $n = N$ being the two edge sites. Note that the translational invariance of the system is broken for open boundary conditions. We numerically find the corresponding quasi energy spectrum. The last figure in the Fig.1 plots the quasi energy spectrum as a function of $\Phi$ when $N=40$, $\gamma=1$, $\lambda=0.4$ and $\omega=2\pi$. The system is topologically trivial (nontrivial) when $\ds{\pi/2<\Phi<3\pi/2}$ (otherwise) since $t_1>t_2$ ($t_1<t_2$) in this region. In the absence of non-Hermitian potential, doubly degenerate zero energy states are the topological edge states. The presence of non-Hermitian term lifts the degeneracy and two symmetrical states appear on both neighboring sides of zero energy line. Let us now study evolution of the edge states. We start with an initial edge state obtained in the limit $\gamma=0$. Then we numerically find the evolution by assuming that $\gamma=1$ and $\omega=2\pi$. The Fig.2.a and Fig.2.b are for topologically nontrivial system with $\Phi=0$ ($t_1=0.4, t_2=1.6$) and topologically trivial system with $\Phi=\pi/2$ ($t_1=1.4 , t_2=0.4$), respectively. The edge state is protected in the former case while it is delocalized in the latter case as expected. As opposed to topological edge states for a Hermitian Hamiltonian, the density of the edge state in our non-Hermitian system oscillates while propagating as can be seen in the Fig.2.a. We note that oscillation of the amplitude of the edge state is the direct result of the quasi energy spectrum of the non-Hermitian Hamiltonian. To this end, we would like to emphasize that we numerically see that complex energy eigenvalues start to appear if the non-Hermitian degree $\gamma$ exceed a critical value at fixed $\omega=2\pi$ (the critical non-Hermitian degree $\gamma_c=3.6$ when $\omega=2\pi$). In this case, the system is topologically unstable. The real part of the spectrum changes dramatically once the complex energy appear in the system.\\
\begin{figure}[t]
\label{fig000}
\includegraphics[width=8.25cm]{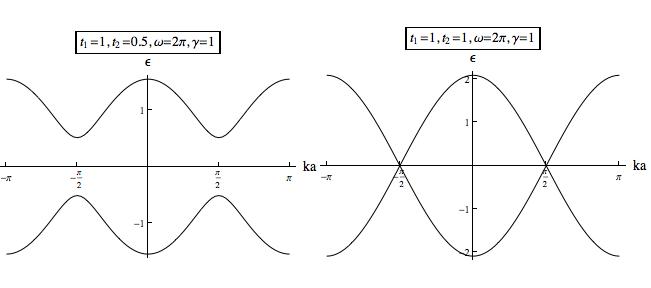}
\includegraphics[width=8.25cm]{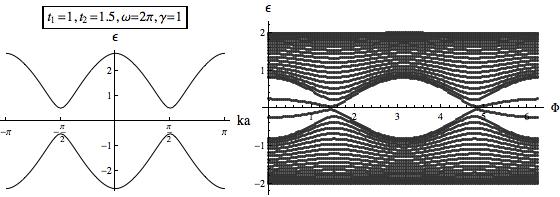}
\caption{The first three figures plot quasi energy spectra of the translationally invariant system for three different coupling coefficient $t_2$. The last figure plots the quasi energy spectrum for the open boundary condition as a function of $\Phi$, where we define for $t_1=1-0.4 \cos(\Phi)$,  $t_2=1+0.4 \cos(\Phi)$. The parameters are given by $\ds{N=40}$, $\omega=2\pi$ and $\gamma=1$.}
\end{figure}
\begin{figure}[t]
\label{fig001}
\includegraphics[width=9cm]{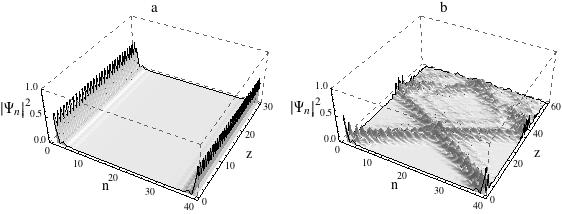}
\caption{The propagation of an initial edge state when $N=40$. The edge state is protected in the topologically nontrivial case (a), while it is delocalized in the topologically trivial case (b). The density would grow without bound when $\omega=0$ while oscillates if the quasi energy spectrum is real when $\omega\neq0$ as can be seen in (a).}
\end{figure}
We have studied the FTI in 1D. Now, our aim is to study the FTI in a 2D system. Consider a 2-dimensional translationally invariant and periodically driven system with the period $T=2\pi/\omega$
\begin{eqnarray}\label{mccnaz}
H=i\gamma\cos(\omega z)\sigma_z+A(\sin(k_xa)\sigma_x+\sin(k_ya)\sigma_y)\nonumber\\
+\left(\Delta+\cos(k_xa)+\cos(k_ya)\right)\sigma_z~~~~~~~~~~~~~~~~
\end{eqnarray}
where $\Delta$ is a real valued constant defined in the interval $\left[-2,2\right]$ and we set $A=1$. The system is supposed to be translationally invariant and wave numbers $k_x$ and $k_y$ are therefore good quantum numbers. The corresponding energy eigenvalues can be found analytically if the system is not periodically modulated, $\omega=0$,
\begin{equation}\label{mccnaz2}
E=\mp(\sin^2k_xx+\sin^2k_yy+(\Delta+\cos {k_x}x+\cos {k_y}y)^2+{\delta}E)^{1/2}
\end{equation}
where ${\delta}E=-\gamma^2+2 i \gamma(\Delta+\cos {k_x}x+\cos {k_y}y)$. In the Hermitian limit, $\gamma=0$, the gap in the spectrum can be closed and reopened by tuning the parameter $\Delta$. If $\Delta=-2$ and $\Delta=2$ the energy gap closes when $k_x = k_y = 0$ and $k_x = k_y = \mp\pi$, respectively. If $\Delta=0$, the energy gap closes when $k_x = 0,k_y = \mp\pi$ and $k_x = \mp\pi,k_y=0$. For all other values of $\Delta$ the spectrum is gapped, and thus nontrivial topological phase exists in the system. In the presence of non-Hermitian potential, $\gamma\neq0$, the system becomes unstable since the system has complex energy eigenvalues. Therefore, no stable topological phase is available for the non-driven case just as for the 1D case considered above. Let us now study the quasi energy spectrum for the periodical system $\omega\neq0$. As in the 1D case, the quasi energy spectrum is real if $\omega$ is large enough for a given $\gamma$. In other words, no complex quasi energy eigenvalues appear during the quantum phase transition. The Fig.3 plots numerically calculated quasi energy spectrum as functions of $k_x$ and $k_y$ for three different values of $\Delta$ at $\omega=6\pi$ and $\gamma=1$. One can see the band gap closing and reopening when we increase $\Delta$ from $-0.5$ to $0.5$. The system is an insulator when $\Delta=\mp0.5$ while the band gap closes at $\Delta=0$. So, we conclude that the two cases with $\Delta=\mp0.5$ are topologically distinct. Since the quasi energy spectrum is real and quantum phase transition occurs during the continuous deformation of the Hamiltonian by varying the parameter $\Delta$, the system is said to be topologically stable. Let us now study edge states. The vacuum is topologically trivial so placing the system in the vacuum leads to the edge states. Suppose the lattice is infinite along $x$ while there are $N$ sites in $y$ axis. The system is subjected to open boundary condition along $y$. Translation invariance holds along $x$ and $k_x$ is still a good quantum number. The last figure in the Fig.3 plots the corresponding quasi energy spectrum as a function of $k_x$ when $\Delta=-0.5$, $\omega=6\pi$ and $\gamma=1$. As can be seen, the spectrum is gapped with gapless edge states crossing between valance and conduction bands. The gapless edge states leave their bulk bands from nearly $k_x=\mp\pi/4$ and cross near $k_x =0$. These states are edge states since their wave functions are concentrated at the edges. However, the states in the bulk band are extended states.  One can notice that the pictures for Hermitian and non-Hermitian cases are similar as long as the spectrum of the non-Hermitian Hamiltonian is real. The real part of the spectrum changes drastically if the imaginary part of it is different from zero. Similar to the 1D case, the quasi energy spectrum becomes complex if $\gamma$ exceed a $\omega$ dependent critical value.\\
To sum up, we have studied $\mathcal{P}$ and $\mathcal{T}$ symmetric non-Hermitian Hamiltonians in 1D and 2D. We consider two band model and show that quantum phase transition occurs in our system. We discuss that topological states are not stable if the system is not periodically modulated. By periodically modulating the non-Hermitian potential, we show that it is possible to get stable Floquet topological phase.
\begin{figure}[t]
\label{fig002}
\includegraphics[width=8cm]{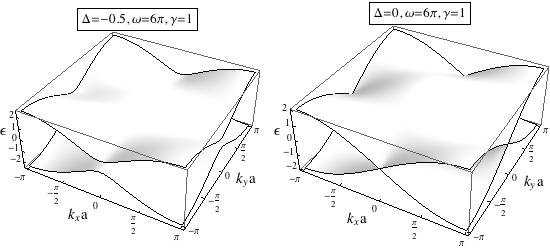}
\includegraphics[width=8.5cm]{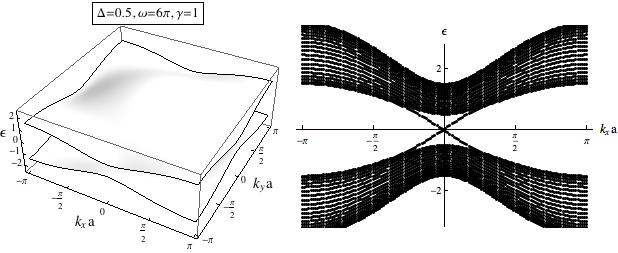}
\caption{The quasi energy spectra for different values of $\Delta$. The gap closes and reopens by tuning $\Delta$. The last figure plots the quasi energy spectrum for the open boundary condition as a function of $k_x$. The parameters are given by $\ds{N=20}$ unit cells wide in the $y$ direction and infinite along the $x$ direction, $\omega=6\pi$ and $\gamma=1$.}
\end{figure}

\end{document}